\documentclass[12pt, draftclsnofoot, onecolumn]{IEEEtran} 

\usepackage{amsmath}
\usepackage{epsfig}
\usepackage{eucal}
\usepackage{amssymb}
\usepackage{bm}
\usepackage{cite}
\usepackage{graphicx}
\usepackage{multirow}
\usepackage{subfigure}
\usepackage{graphicx,color,psfrag}
\usepackage{multirow}

\usepackage{array, tabularx, colortbl, color, threeparttable}
\usepackage[table]{xcolor}
\usepackage{psfrag, pstricks}
\usepackage{booktabs}

\newtheorem{lem}{Lemma}
\newtheorem{thm}{Theorem}
\newtheorem{cor}{Corollary}

\newcommand{\paren}[1]{\left(#1\right)}
\newcommand{\sqparen}[1]{\left[#1\right]}
\newcommand{\brparen}[1]{\left\{#1\right\}}
\newcommand{\abs}[1]{\left| #1\right|}
\newcommand{\field}[1]{\ensuremath{\mathbb{#1}}}
\newcommand{\R}{\ensuremath{\field{R}}} 
\newcommand{\I}[1]{\ensuremath{\mathsf{1}_{\left\{#1\right\}}}} 
\newcommand{\Inb}[1]{\ensuremath{\mathsf{1}_{#1}}} 
\newcommand{\ra}{\ensuremath{\rightarrow}} 
\newcommand{\PR}[1]{\ensuremath{\mathsf{Pr}\left\{#1\right\}}} 
\newcommand{\EW}{\ensuremath{\mathsf{E}}} 
\newcommand{\ES}[1]{\ensuremath{\mathsf{E}\left[#1 \right]}} 
\newcommand{\V}[1]{\ensuremath{\mathsf{Var}\left(#1 \right)}} 
\newcommand{\e}[1]{\ensuremath{{\rm e}^{#1}}} 

\newcommand{\Ceil}[1]{\ensuremath{\left\lceil #1\right\rceil}}

\newcommand{\BO}[1]{\ensuremath{O\paren{#1}}} 
\newcommand{\vecbold}[1]{\ensuremath{\boldsymbol{#1}}}

\renewcommand{\vec}[1]{\ensuremath{\boldsymbol{#1}}} 

\begin{document}
%
\title{Gaussian Approximation for the Downlink Interference in Heterogeneous Cellular Networks}

\author{Serkan Ak,~\IEEEmembership{Student Member,~IEEE}, Hazer Inaltekin,~\IEEEmembership{Member,~IEEE}, \\ H. Vincent Poor,~\IEEEmembership{Fellow,~IEEE} 

 } 

\markboth{}{Ak \MakeLowercase{\textit{et al.}}: Gaussian Approximation for the Downlink Interference in Heterogeneous Cellular Networks}


\maketitle
\begin{abstract}
This paper derives Gaussian approximation bounds for the standardized aggregate wireless interference (AWI) in the downlink of {\em dense} $K$-tier heterogeneous cellular networks when base stations in each tier are distributed over the plane according to a (possibly non-homogeneous) Poisson process. The proposed methodology is general enough to account for general bounded path-loss models and fading statistics. The deviations of the distribution of the standardized AWI from the standard normal distribution are measured in terms of the Kolmogorov-Smirnov distance. An explicit expression bounding the Kolmogorov-Smirnov distance between these two distributions is obtained as a function of a broad range of network parameters such as per-tier transmission power levels, base station locations, fading statistics and the path-loss model. A simulation study is performed to corroborate the analytical results. In particular, a good statistical match between the standardized AWI distribution and its normal approximation occurs even for moderately dense heterogeneous cellular networks. These results are expected to have important ramifications on the characterization of performance upper and lower bounds for emerging 5G network architectures. 

\end{abstract}


\IEEEpeerreviewmaketitle
\normalsize
\section{Introduction}\label{Introduction}
The next generation of wireless networks is envisioned to be more heterogenous and denser in order to meet high capacity demands from mobile users \cite{Qualcomm1,Ghosh12,Hwang13}.
Therefore, characterization and mitigation of aggregate wireless interference (AWI) appear to be a more pronounced design bottleneck against meeting such high data rate demands in heterogenous cellular networks (HCNs), e.g., see \cite{Seymour14} and \cite{Hossain14}. However, even for traditional macro cell deployments, computation of the exact AWI distribution is a very challenging task that usually does not result in closed form expressions \cite{HG09, Andrews11}.  This motivates us in the current paper to search for a structure in the distribution of AWI for the downlink of a $K$-tier HCN that will lead to simplifications in performance characterization and network design. 

The early work in the literature focusing on approximating the distribution of AWI in wireless networks includes \cite{IH98, CH01, WA06}. These papers considered traditional single-tier macro cell deployments and obtained various approximations on the distribution of AWI using LePage series \cite{IH98}, Edgeworth expansion \cite{CH01} and geometrical considerations \cite{WA06}. More recently, Berry-Esseen types of bounds were obtained in \cite{AY10, Inaltekin12}, but again by considering only single-tier wireless networks. The related work also includes those papers \cite{Heath13, Mukherjee12} on the analysis of interference and signal-to-interference-plus-noise ratio (SINR) in the downlink of $K$-tier HCNs. In \cite{Heath13}, the authors investigated a Gamma distribution approximation for the distribution of AWI  clogging a fixed-size cell with a guard zone and a dominant interferer. In \cite{Mukherjee12}, the author derived the downlink SINR distribution for $K$-tier HCNs by assuming the classical unbounded path-loss model, Rayleigh faded wireless links and the nearest base-station (BS) association rule.     

In this paper, we examine the problem of Gaussian approximation for the standardized (i.e., centered and normalized) AWI in the downlink of a {\em dense} $K$-tier HCN, where the network tiers are differentiated from each other in terms of transmission power levels, spatial BS distribution and RF signal propagation characteristics.  In particular, the underlaying spatial stochastic processes determining the BS locations in each tier are assumed to be Poisson but not necessarily homogenous. The signal power attenuation due to path-loss is modeled through a general bounded and power-law decaying path-loss function, which can vary from one tier to another.  Fading and shadowing are also accounted for in the employed signal propagation model without assuming any specific distribution functions for these other random wireless channel dynamics. 

Measuring the distance between the standardized downlink AWI and normal distributions by means of Kolmogorov-Smirnov statistic, we obtain an analytical expression for deviations between them. This is the main contribution of the present paper. Briefly, the stated distance consists of two parts: (i) a scaling coefficient, multiplied with (ii) a positive function $c(x)$ with $x \in \R$ being the point at which we want to estimate the value of the standardized AWI distribution. The scaling coefficient depends on various network parameters at each tier such as transmission powers, BS distribution and signal propagation characteristics. An important property of the scaling coefficient is its monotonically decaying nature to zero with denser deployments of BSs per tier. On the other hand, the function $c(x)$ is {\em uniformly bounded} by a small constant and approaches zero for large absolute values of $x$ at a rate $\abs{x}^{-3}$, which makes the derived bounds on the tails of the standardized downlink AWI distribution tight even for sparsely deployed HCNs. These results are formally given in Theorem \ref{Theorem 1}.                   

The above stated contributions in this paper differ from the previous work in several important aspects. When compared to \cite{IH98, CH01, WA06, AY10, Inaltekin12}, this paper extends the previous known results approximating AWI distribution for macro cell deployments to more heterogenous and complex wireless communication environments. In particular, functional dependencies among different tiers to approximate the AWI distribution in the downlink of a HCN are clearly identified.  When compared with the results reported in \cite{Heath13, Mukherjee12}, our network set-up is much richer, allowing non-homogenous Poisson point processes (PPP) for BS locations and general signal propagation models including fading and shadowing.

\section{System Model}\label{System Model}
In this section, we will introduce the details of the studied downlink model in a $K$-tier cellular topology, the details of the spatial processes determining BS locations and the signal propagation characteristics.  
\subsection{The Downlink Model in a $K$-Tier Cellular Topology} 
We consider an overlay $K$-tier HCN in which the BSs in all tiers are fully-loaded (i.e., no empty queues) and access to the same communication resources both in time and frequency.  The BSs in different tiers are differentiated mainly on the basis of their transmission powers, with $P_k > 0$ being the transmission power of a tier-$k$ BS for $k = 1, \ldots, K$.  As is standard in stochastic geometric modeling, it is assumed that BSs  are distributed over the plane according to a PPP (possibly non-homogeneous) with differing spatial density among the tiers.  Further, the signal propagation characteristics (including both large-scale path-loss and small-scale fading effects) also vary from one tier to another. The details of BS location processes and signal propagation are elaborated below.  

We place a test user at an arbitrary point $\vecbold{x}^{(o)} = \paren{x^{(o)}_1, x^{(o)}_2} \in \R^2$ and consider signals coming from all the BSs in all tiers as the {\em downlink} AWI experienced by this test user.  Since we focus on the downlink analysis, we assume that the uplink and downlink do not share any common communication resources.  Therefore, the uplink interference can be ignored for the analysis of downlink AWI.  This setting is general enough to understand the effects of various network parameters such as transmission powers and BS intensity in each tier on the distribution of the AWI seen by the test user.

\subsection{BS Location Processes}
The BS locations in tier-$k$, $k=1, \ldots, K$, independently form a spatial planar PPP $\Phi _{\Lambda^{\paren{k}}}$, where $\Lambda^{(k)}$ represents the {\em mean measure} (alternatively called: intensity measure or spatial density) of the $k$th tier BSs.  We do not assume any specific functional form for $\Lambda^{(k)}$ and hence do not restrict our attention only to homogenous PPPs. For each (Borel) subset $\mathcal{A}$ of $\R^2$, $\Lambda^{(k)}\paren{\mathcal{A}}$ gives us the average number of BSs lying in $\mathcal{A}$. We will assume that $\Lambda^{(k)}$ is {\em locally finite} i.e., $\Lambda^{(k)}\paren{\mathcal{A}} < \infty$ for all bounded subsets $\mathcal{A}$ of $\R^2$, and $\Lambda^{(k)}\paren{\R^2} = \infty$, i.e., there is an infinite population of tier-$k$ BSs scattered all around in $\R^2$.  For the whole HCN, the aggregate BS location process, which is the superposition of all individual position processes, is denoted by ${\Phi _\Lambda } = \bigcup\nolimits_{k = 1}^K {{\Phi _{{\Lambda^{( k)}}}}}$. Henceforth, when we refer to an interfering BS (without specifying its tier) in the sequel, we write $\vecbold{X} \in \Phi_\Lambda$ to represent its location. 

For mathematical convenience, we also express $\Phi _{\Lambda^{\paren{k}}}$ as a discrete sum of Dirac measures as $\Phi _{\Lambda^{\paren{k}}} = \sum_{j \geq 1} \delta_{\vecbold{X}_j^{(k)}}$, where $\delta_{\vecbold{X}_j^{(k)}}\paren{\mathcal{A}} = 1$ if $\vecbold{X}_j^{(k)} \in \mathcal{A} \subseteq \R^2$, and zero otherwise.  The level of AWI at $\vecbold{x}^{(o)}$ from tier-$k$ BSs depends critically on the distances between the points of $\Phi _{\Lambda^{\paren{k}}}$ and $\vecbold{x}^{(o)}$.  It is well-known from the theory of Poisson processes that the transformed process $ \sum_{j \geq 1} \delta_{T\paren{\vecbold{X}_j^{(k)}}}$ is still Poisson (on the positive real line) with mean measure given by $\Lambda^{(k)} \circ T^{-1}$, where $T\paren{\vecbold{x}} = \left\| \vecbold{x} - \vecbold{x}^{(o)} \right\|_2 = \sqrt{\paren{x_1 - x^{(o)}_1}^2 +  \paren{x_2 - x^{(o)}_2}^2}$ and $T^{-1}\paren{\mathcal{A}} = \brparen{\vecbold{x} \in \R^2: T\paren{\vecbold{x}} \in \mathcal{A}}$ for all $\mathcal{A} \subseteq \R$ \cite{Kingman93}. We will assume that $\Lambda^{(k)} \circ T^{-1}$ has a density in the form $\Lambda^{(k)} \circ T^{-1}\paren{\mathcal{A}} = \lambda_k \int_{\mathcal{A}} \mu_k(t) dt$.  Here, $\lambda_k$ is a modeling parameter pertaining to the $k$th tier, which can be interpreted as the {\em BS intensity parameter}, that will enable us to control the average number of tier-$k$ BSs whose distances from $\vecbold{x}^{(o)}$ belong to $\mathcal{A}$ and interfere with the signal reception at the test user.   

\subsection{Signal Propagation and Interference Power}  
 We model the large scale signal attenuation for tier-$k$, $k=1, \ldots, K$, by a {\em bounded} monotone non-increasing path-loss function $G_k: [0, \infty) \mapsto [0, \infty)$. $G_k$ asymptotically decays to zero at least as fast as $t^{-\alpha_k}$ for some path-loss exponent $\alpha_k > 2$.  To ensure the finiteness of AWI at the test user, we require the relationship ${\mu}_{k}(t) = \BO{t^{\alpha_k - 1 - \epsilon}}$ as $t \to \infty$ to hold for some $\epsilon > 0$.

The fading (power) coefficient for the wireless link between a BS located at point $\vecbold{X} \in \Phi_{\Lambda}$ and the test user is denoted by $H_{\vecbold{X}}$.\footnote{For simplicity, we only assign a {\em single} fading coefficient to each BS. In reality, it is expected that the channels between a BS and all potential receivers (intended or unintended) experience different (and possibly independent) fading processes. Our simplified notation does not cause any ambiguity here since we focus on the total interference power at a given arbitrary position in $\mathbb{R}^{2}$ in the remainder of the paper.} The fading coefficients $\brparen{H_{\vecbold{X}}}_{\vecbold{X} \in \Phi_\Lambda}$ form a collection of independent random variables (also independent of ${\Phi _\Lambda }$), with those belonging to the same tier, say tier-$k$, having a common probability distribution with density $q_k(h), h \geq 0$.  The first, second and third order moments of fading coefficients are assumed to be finite, and are denoted by $m^{(k)}_{H}$, $m^{(k)}_{H^2}$ and $m^{(k)}_{H^3}$, respectively, for tier-$k$.  We note that this signal propagation model is general enough that $H_{\vecbold{X}}$'s could also be thought to incorporate {\em shadow fading} effects due to blocking of signals by large obstacles existing in the communication environment, although we do not model such random factors explicitly and separately in this paper.

Considering all the signal impairments due to fading and path-loss, we can write the interference power seen by the test user from a tier-$k$ BS located at $\vecbold{X}^{(k)} \in \Phi_{\Lambda^{(k)}}$ as ${P_k}H_{\vecbold{X}^{(k)}}G_k\paren{T\paren{\vecbold{X}^{(k)}}}$.  Hence, the level of AWI at $\vecbold{x}^{(o)}$ is equal to $I_{\vecbold{\lambda}} = \sum_{k = 1}^K \sum_{\vecbold{X}^{(k)} \in \Phi _{{\Lambda ^{(k)}}}} {P_k}{H_{\vecbold{X}^{(k)}}}G_k\paren{T\paren{\vecbold{X}^{(k)}}}$, where $\vecbold{\lambda} = \sqparen{\lambda_1, \ldots, \lambda_K}^\top$. This parametrization of AWI is chosen to emphasize the dependence of its distribution to the BS intensity parameter $\lambda_k$ of each tier. $I_{\vecbold{\lambda}}$ is a random function of BS configurations and fading states. In the next section, we will show that the distribution of $I_{\vecbold{\lambda}}$ can be approximated by a Gaussian distribution.

\section{Gaussian Approximation for the AWI Distribution} \label{Gaussian Approximation for Interference Distribution}
In this section, we will establish the Gaussian approximation bounds for the distribution of the standardized AWI in the downlink of a HCN.  These bounds will clearly show the functional dependence between the downlink AWI distribution and a broad range of network parameters such as transmission power levels, BS distribution over the plane and signal propagation characteristics in each tier. We will also specialize these approximation results to the commonly used homogenous PPPs at the end of this section. Most of the proofs are relegated to appendices for the sake of fluency of the paper. Hence, we focus on the main engineering and design implications of these results for emerging 5G networks in the remainder of the paper.  

\begin{thm} \label{Theorem 1} 
For all $x \in \mathbb{R}$,
\begin{eqnarray}
\abs{\PR{\frac{I_{\vecbold{\lambda}} - \ES{I_{\vecbold{\lambda}}}}{\sqrt{\V{I_{\vecbold{\lambda}}}}} \leq x} - \Psi(x)} \leq \Xi \cdot c(x),
\end{eqnarray}
where $\Xi = \sum_{k=1}^K \frac{\lambda_k P_k^3 m_{H^3}^{(k)} \int_0^\infty G_k^3(t) \mu_k(t) dt}{\paren{\sum_{k=1}^K \lambda_k P_k^2 m_{H^2}^{(k)} \int_0^\infty G_k^2(t) \mu_k(t) dt}^\frac32}$, $c(x) = \min\paren{0.4785, \frac{31.935}{1 + \abs{x}^3}}$ and $\Psi(x) = \frac{1}{\sqrt{2\pi}}\int_{-\infty}^x \e{- \frac{t^2}{2}} dt$, which is the standard normal cumulative distribution function (CDF). 
\end{thm}
\begin{IEEEproof}
Please see Appendix \ref{Appendix:Theorem_1}. 
\end{IEEEproof}

Measuring the distance by means of Kolmogorov-Smirnov statistic, Theorem \ref{Theorem 1} provides us with an explicit expression for the deviations between the standardized AWI and normal distributions.  Several important remarks about this result are in order. The scaling coefficient $\Xi$ appearing in Theorem  \ref{Theorem 1} is linked to the main network parameters such as transmission power levels, distribution of BSs over the plane and signal propagation characteristics.  Starting with the BS intensity parameters $\lambda_k$, $k = 1, \ldots, K$, we observe that the rate of growth of the expression appearing in the denominator of $\Xi$ is half an order larger than that of the expression appearing in the numerator of $\Xi$ as a function of $\lambda_k$. This observation implies that the derived Gaussian approximation becomes tighter for denser deployments of HCNs.  A formal statement of this result is given in the following lemma. 
\begin{lem} \label{Lemma: Convergence Rate}
The scaling coefficient $\Xi$ appearing in the Gaussian approximation result in Theorem \ref{Theorem 1} is bounded above by $\Xi \leq \frac{\delta}{\sqrt{\| \vecbold{\lambda} \|_2}}$ for some finite positive constant $\delta$.   
\end{lem}       
\begin{IEEEproof}
Let $a_k = P_k^3 m_{H^3}^{(k)} \int_0^\infty G_k^3(t) \mu_k(t) dt$ and $b_k = P_k^2 m_{H^2}^{(k)} \int_0^\infty G_k^2(t) \mu_k(t) dt$. Then, 
\begin{eqnarray}
\Xi &=& \frac{\sum_{k=1}^K a_k \lambda_k}{\paren{\sum_{k=1}^K \lambda_k b_k}^\frac32} \leq \frac{\| \vecbold{\lambda}\|_2 \| \vecbold{a} \|_2}{\paren{\sum_{k=1}^K \lambda_k b_k}^\frac32} \nonumber
\end{eqnarray}
due to Cauchy-Schwarz inequality. Further, we can lower-bound the sum in the denominator above as
\begin{eqnarray}
\paren{\sum_{k=1}^K \lambda_k b_k}^\frac32 \geq \paren{\min_{1 \leq k \leq K} b_k \sum_{k=1}^K \abs{\lambda_k}}^\frac32 \geq \epsilon \paren{\| \vecbold{\lambda} \|_2}^\frac32, \nonumber
\end{eqnarray}
where the last inequality follows from the equivalence of all the norms in finite dimensional vector spaces. Combining these two inequalities, we conclude the proof. 
\end{IEEEproof}

Following a similar approach, we can also see that changing transmission powers is not as effective as changing BS intensity parameters to improve the Gaussian approximation bound in Theorem \ref{Theorem 1}. This is expected since the power levels are assumed to be deterministic (i.e., no power control is exercised) and therefore they do not really add to the randomness coming from the underlying spatial BS distribution over the plane and the path-loss plus fading characteristics modulating transmitted signals.  

Another important observation we have in regards to the combined effect of the selection of transmission powers per tier and the moments of fading processes in each tier on the Gaussian approximation result in Theorem \ref{Theorem 1} is that our approximation bounds benefit from the fading distributions with restricted dynamic ranges and the alignment of received AWI powers due to fading and path-loss components.  This observation is made rigorous through the following lemma.    
\begin{lem} \label{Lemma: Effect of Fading}
Let $a_k = \lambda_k \int_0^\infty G_k^3(t) \mu_k(t) dt$, $b_k = \lambda_k \int_0^\infty G_k^2(t) \mu_k(t) dt$ and $c_k = P_k^2 m_{H^2}^{(k)}$.  Then, the scaling coefficient $\Xi$ appearing in the Gaussian approximation result in Theorem \ref{Theorem 1} is bounded below by 
$$ \Xi \geq \paren{\frac{1}{\|\vecbold{c}\|_2 \| \vecbold{b} \|_2}}^\frac32 \sum_{k=1}^K a_k c_k^\frac32,$$
with equality achieved if fading processes in all tiers are deterministic and the vectors $\vecbold{b} = \sqparen{b_1, \ldots, b_K}^\top$ and $\vecbold{c} = \sqparen{c_1, \ldots, c_K}^\top$ are parallel.  
\end{lem}
\begin{IEEEproof}
Using $a_k, b_k$ and $c_k$ introduced above, we can write a lower bound for $\Xi$ as
\begin{eqnarray*}
\Xi = \frac{\sum_{k=1}^K a_k P_k^3 m_{H^3}^{(k)}}{\paren{\sum_{k=1}^K b_k c_k}^\frac32} &=&  \frac{\sum_{k=1}^K a_k P_k^3 m_{H^3}^{(k)}}{\paren{\| \vecbold{c} \|_2}^\frac32 \paren{\sum_{k=1}^K b_k \frac{c_k}{\| \vecbold{c}\|_2}}^\frac32} \\
&\geq& \paren{\frac{1}{\| \vecbold{c} \|_2 \| \vecbold{b} \|_2}}^\frac32 \sum_{k=1}^K a_k P_k^3 m_{H^3}^{(k)}.
\end{eqnarray*}
Using Jensen's inequality, we also have $m_{H^3}^{(k)} \geq \paren{m_{H^2}^{(k)}}^\frac32$. Using this lower bound on $m_{H^3}^{(k)}$ in the above expression, we finally have $ \Xi \geq \paren{\frac{1}{\| \vecbold{c} \|_2 \| \vecbold{b} \|_2}}^\frac32 \sum_{k=1}^K a_k c_k^\frac32$. 
\end{IEEEproof}

In addition to the above fundamental properties of the scaling coefficient $\Xi$, it is also worthwhile to mention that the Gaussian approximation bound derived in Theorem \ref{Theorem 1} is a combination of two different types of Berry-Esseen bounds embedded in the function $c(x)$.  One of these bounds is a {\em uniform} bound that helps us to estimate the standardized AWI distribution uniformly as
$$ \abs{\PR{\frac{I_{\vecbold{\lambda}} - \ES{I_{\vecbold{\lambda}}}}{\sqrt{\V{I_{\vecbold{\lambda}}}}} \leq x} - \Psi(x)} \leq \Xi \cdot 0.4785. $$
On the other hand, the other one is a {\em non-uniform} bound that helps us to estimate the {\em tails} of the standardized AWI distribution as
$$ \abs{\PR{\frac{I_{\vecbold{\lambda}} - \ES{I_{\vecbold{\lambda}}}}{\sqrt{\V{I_{\vecbold{\lambda}}}}} \leq x} - \Psi(x)} \leq \Xi \cdot \frac{31.935}{1+\abs{x}^3}$$ 
and decays to zero as a third order inverse power law.

Up to now, we considered general PPPs for the distribution of BSs in each tier. One simplifying assumption in the literature is to assume that PPPs determining the locations of BSs are homogenous. In this case, $\mu_k(t)$ for all tiers is given by $\mu_k(t) = 2\pi t \I{t \geq 0}$, where $\I{\cdot}$ is the indicator function.  Using this expression for $\mu_k(t)$ in Theorem \ref{Theorem 1}, we obtain the following approximation result for the distribution of AWI when all BSs are homogeneously distributed over the plane according to a PPP with differing BS intensity parameters $\lambda_k$ from tier to tier.   
 
\begin{thm} \label{Theorem 2} 
Assume that $\Phi_{\Lambda^{(k)}} $ is a homogeneous PPP with a mean measure given $\Lambda^{(k)}\paren{\mathcal{A}}  = \lambda_k \cdot \mbox{area}\paren{\mathcal{A}}$. Then, for all $x \in \R$,
\begin{eqnarray}
\abs{\PR{\frac{I_{\vecbold{\lambda}} - \ES{I_{\vecbold{\lambda}}}}{\sqrt{\V{I_{\vecbold{\lambda}}}} } \le x} - \Psi(x)} \leq \Xi \cdot c(x),
\end{eqnarray}
where $\Xi = \frac{1}{\sqrt{2\pi}} \sum_{k=1}^K \frac{\lambda_k P_k^3 m_{H^3}^{(k)} \int_0^\infty G_k^3(t) t dt}{\paren{\sum_{k=1}^K \lambda_k P_k^2 m_{H^2}^{(k)} \int_0^\infty G_k^2(t) t dt}^\frac32}$, $c(x) = \min\paren{0.4785, \frac{31.935}{1+\abs{x}^3}}$ and $\Psi(x) = \frac{1}{\sqrt{2\pi}}\int_{-\infty}^x \e{- \frac{t^2}{2}} dt$, which is the standard normal CDF.    
\end{thm}
\begin{IEEEproof}
The proof follows from Theorem \ref{Theorem 1} by replacing $\mu_k(t)$ with $2\pi t \I{t \geq 0}$.
\end{IEEEproof}

When all network parameters are assumed to be the same, i.e., the same transmission power levels, fading distributions and BS distributions for all tiers, the HCN in question collapses to a single tier network. In this case, the Gaussian approximation result is given below.  
\begin{cor}\label{Corollary_1}
Assume $P_k = P$, $\mu_k(t) = 2\pi t \I{t \geq 0}$, ${G_k}\left( t \right) = G\left( t \right)$, $\lambda_k = \lambda$, $m_{H^2}^{(k)} = m_{H^2}$ and $m_{H^3}^{(k)} = m_{H^3}$ for all $k=1, \ldots, K$. Then, for all $x \in \R$, we have
$$ \abs{\PR{\frac{I_{\vecbold{\lambda}} - \ES{I_{\vecbold{\lambda}}}}{\sqrt{\V{I_{\vecbold{\lambda}}}}} \le x} - \Psi(x)} \leq \Xi \cdot c(x), $$
where $\Xi = \frac{1}{\sqrt{2 \pi}} \frac{1}{\sqrt{K \lambda}} \frac{m_{H^3}}{\paren{m_{H^2}}^\frac32} \frac{\int_0^\infty G^3(t) t dt}{\paren{\int_0^\infty G^2(t) t dt}^\frac32}$, and $c(x)$ and $\Psi(x)$ are as given in Theorem \ref{Theorem 1}. 
\end{cor}

We note that this is the same result obtained in \cite{Inaltekin12} as a special case of the network model studied in this paper.

\section{Numerical Examples}\label{Verification of Gaussian Approximation Through Simulations}
\begin{figure*}[!t]
\begin{minipage}[b]{0.45\linewidth} 
\centering
\includegraphics[width=3.5in]{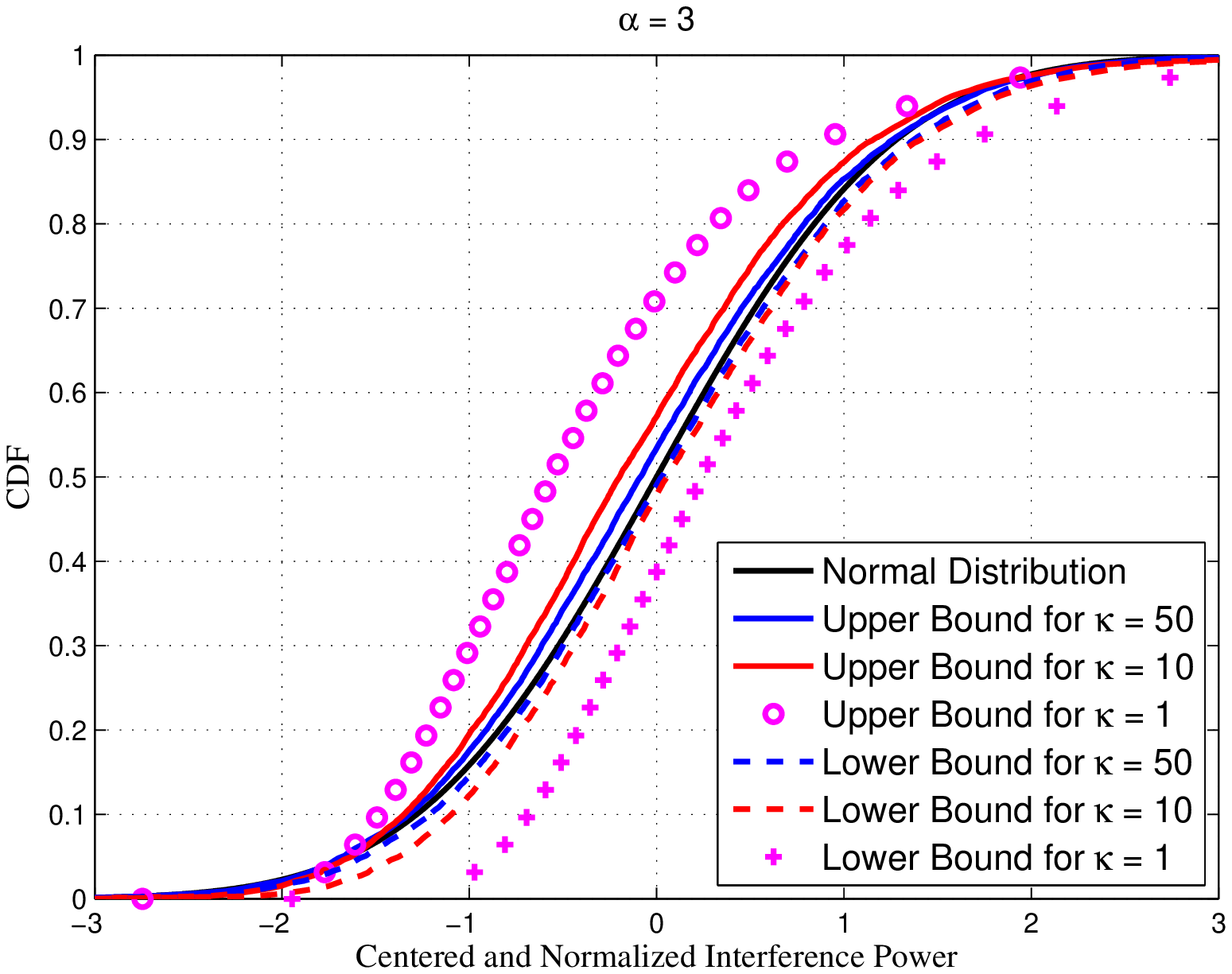}
\end{minipage}
\hspace{0.6cm} 
\begin{minipage}[b]{0.45\linewidth}
\centering
\includegraphics[width=3.5in]{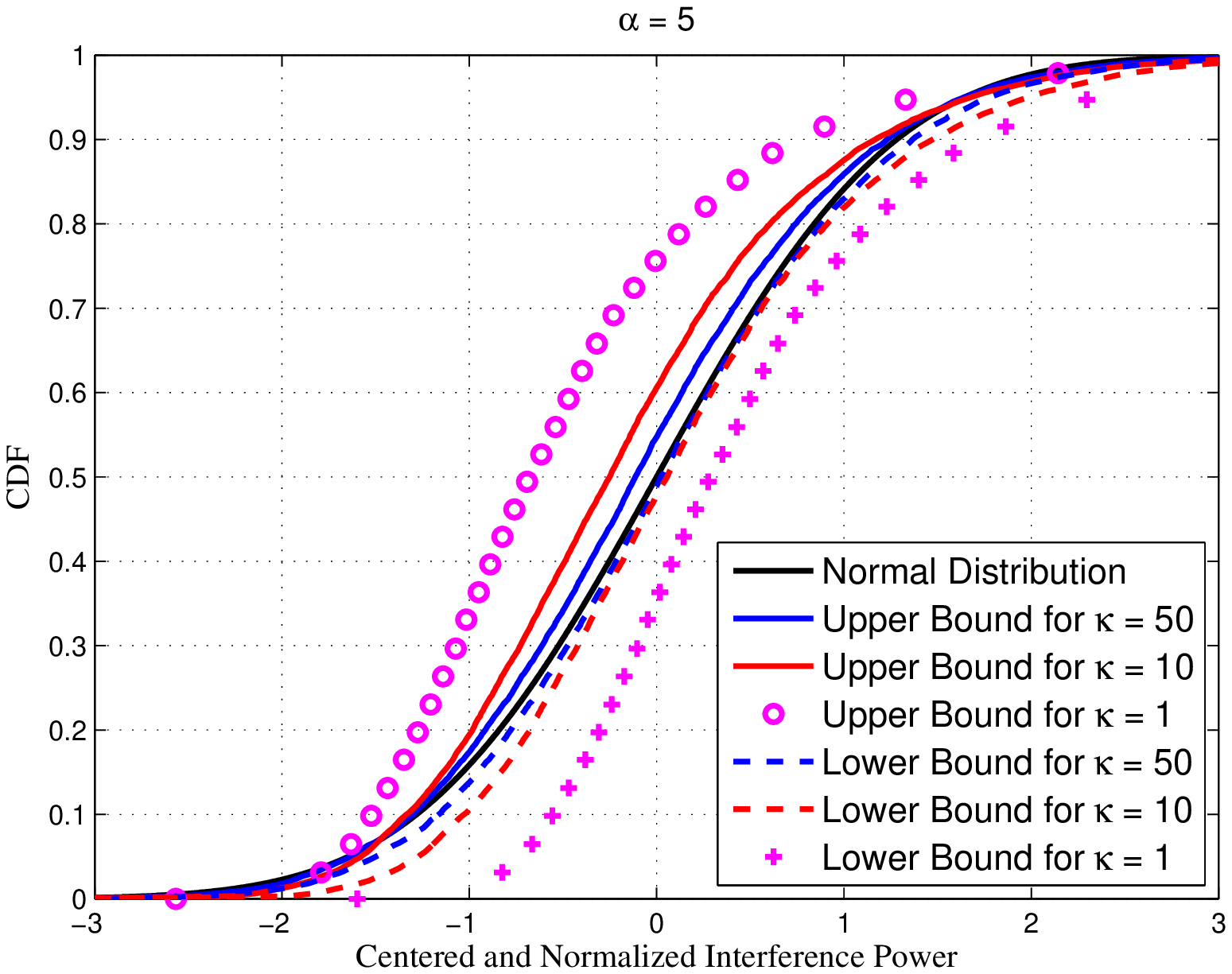}
\end{minipage}
\\
\begin{minipage}[b]{0.45\linewidth} 
\centering
\includegraphics[width=3.5in]{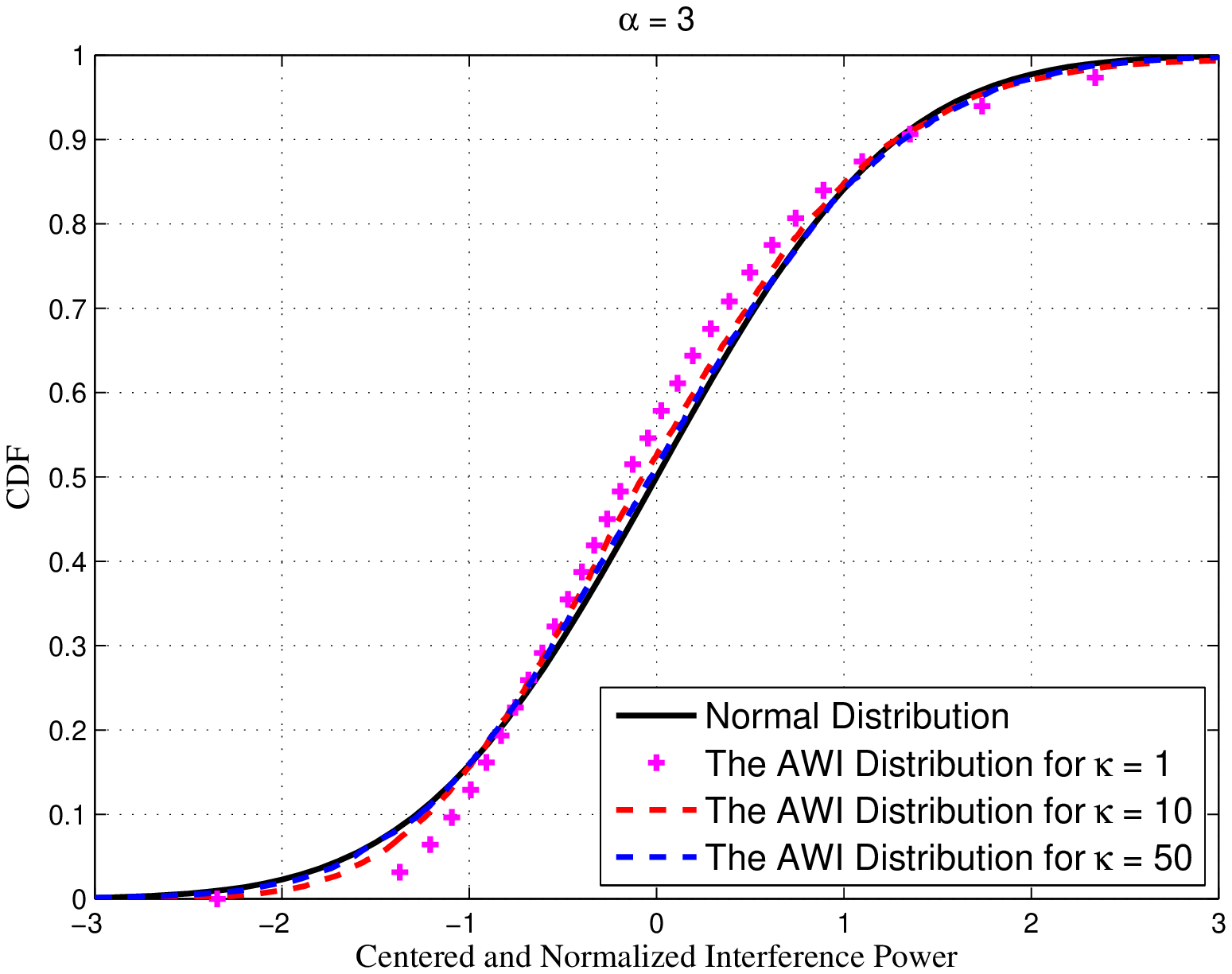}
\end{minipage}
\hspace{0.6cm} 
\begin{minipage}[b]{0.45\linewidth}
\centering
\includegraphics[width=3.5in]{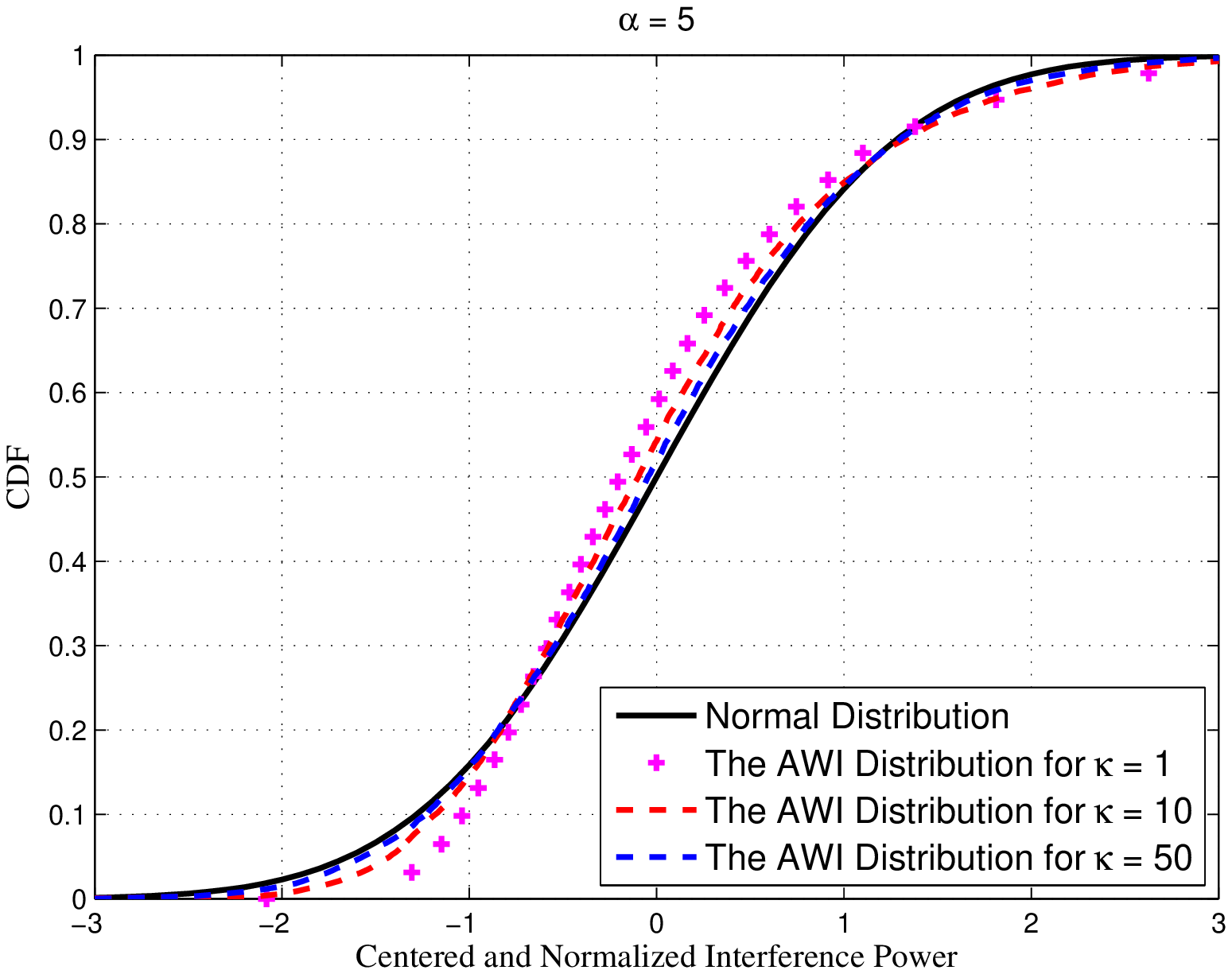}
\end{minipage}
\caption{Gaussian approximation bounds for the standardized AWI CDFs (upper figures). Comparison of the simulated standardized AWI CDFs with the standard normal CDF (lower figures). Rayleigh fading with unit mean power is assumed.} 
\label{fig1}
\end{figure*}
In this section, we will illustrate the analytical Gaussian approximation results derived for the standardized AWI distribution in Section \ref{Gaussian Approximation for Interference Distribution} for a specific three-tier HCN scenario.  To this end, we will assume the same path-loss model $G_k\paren{t} = \frac{1}{{1 + {t^\alpha }}}$ for all tiers with various values of $\alpha > 2$. Similar conclusions continue to hold for other path-loss models.  The BSs in each tier are distributed over the plane according to a homogeneous PPP, with BS intensity parameters given by $\lambda _1 = 0.1\kappa$, $\lambda _2 = \kappa$ and $\lambda _3 = 5\kappa$.  Here, $\kappa$ is our control parameter to control the average number of BSs interfering with the signal reception at the test user.  The test user is assumed be located at the origin without loss of any generality since we focus only on homogenous PPPs in this numerical study.  The random fading coefficients in all tiers are assumed to be i.i.d. random variables, drawn from a Rayleigh distribution with unit mean power gain. Our results are qualitatively the same for other fading distributions such as Nakagami and Rician fading distributions.  The transmission power levels are set as $P_1 = 4 P_2 = 16 P_3$, where $P_2$ is assumed to be unity. 

In the upper figures of Fig. \ref{fig1}, we present the upper and lower bounds for the Kolmorov-Smirnov distance between the standardized AWI and normal distributions, i.e., we plot the expressions $\Psi(x) + \Xi \cdot c(x) $ and $\Psi(x) - \Xi \cdot c(x)$ appearing in Theorem \ref{Theorem 1}, with a variety of $\kappa$ values.  Two different regimes are apparent in these figure.  For the moderate values at which we want to estimate the CDF of standardized AWI, i.e., $\PR{\frac{I_{\vecbold{\lambda}} - \ES{I_{\vecbold{\lambda}}}}{\sqrt{\V{I_{\vecbold{\lambda}}}}} \leq x}$ with moderate $x$ values, our uniform Berry-Esseen bound, which is $\Xi \cdot 0.4785$, provides better estimates for the AWI distribution. On the other hand, for absolute values larger than $3.4$ at which we want to estimate the CDF of standardized AWI, i.e.,  $\PR{\frac{I_{\vecbold{\lambda}} - \ES{I_{\vecbold{\lambda}}}}{\sqrt{\V{I_{\vecbold{\lambda}}}}} \leq x}$ with $\abs{x}$ larger than $3.4$, our non-uniform Berry-Esseen bound, which is $\Xi \cdot \frac{31.935}{1+\abs{x}^3}$, is tighter. These figures also clearly demonstrate the effect of BS intensity parameters $\lambda_k$ on our Gaussian approximation bounds. As suggested by Lemma \ref{Lemma: Convergence Rate}, the Kolmogorov-Smirnov distance between the standardized AWI and normal distributions approach the zero at a rate $\frac{1}{\sqrt{\| \vecbold{\lambda} \|_2}}$. Further, even if all BS intensity parameters are fixed, the distance between the upper and lower bounds in Theorem \ref{Theorem 1} disappears at a rate $O\paren{\abs{x}^{-3}}$ as $\abs{x} \to \infty$ due to the non-uniform bound. 

When we compare upper lefthand side and righthand side figures in Fig. \ref{fig1}, we observe a better convergence behavior for smaller values of the path-loss exponent $\alpha$. This is due to the path-loss model dependent constants appearing in Theorem \ref{Theorem 1}. For this particular choice of path-loss model and BS distribution over the plane, our approximation results benefit from small values of path-loss exponent, although the difference between them becomes negligible for moderate to high values of $\kappa$. 

We also performed Monte-Carlo simulations to compare simulated standardized AWI distributions with the normal distribution for  $10^4$ random BS configurations. The lower figures in Fig. \ref{fig1} provides further numerical evidence for the Gaussian approximation of AWI in HCNs. Surprisingly, there is a good match between the simulated standardized AWI distribution and the standard normal CDF even for sparsely populated HCNs, i.e., $\kappa = 1$.

\section{Conclusions}\label{Conclusion}
In this paper, we have investigated the Gaussian approximation for the AWI distribution in the downlink of HCNs under a general set-up. Analytical bounds measuring the Kolmogorov-Smirnov distance between these two distributions have been obtained. The derived Gaussian approximation bounds have also been illustrated numerically through simulation of a particular three-tier HCN scenario. A good statistical fit between the simulated (centralized and normalized) AWI distribution and the standard normal distribution has been observed even for moderate values of BS intensities.

\vspace{-0.1cm}

\vspace{-0.3cm}
\appendices
\section{Auxiliary Lemmas for the Proof of Theorem \ref{Theorem 1}} \label{Preliminary}
In this appendix, we will provide five lemmas to construct the proof of Theorem \ref{Theorem 1} in the next appendix.  
We start our analysis by showing that AWI has a probability non-degenerate distribution.  By using Laplace functionals of Poisson processes (refer to \cite{Kingman93} for details), we can find the Laplace transform for $I_{\vec{\lambda}}$ as follows:
$$ {{\mathcal{L}}_{{I_{\vec{\lambda}}}}}\left( s \right) = \ES{ {{e^{ - s{I_{\vec{\lambda}} }}}} } = \prod\limits_{k = 1}^K {\exp \left( { - {\lambda _k}\int_0^\infty  {\int_0^\infty  {\left( {1 - {e^{ - s{{P}_{k}}hG_{k}\left( t \right)}}} \right)} } {{\mu}_{k}}\left( t \right){q_{k}}\left( h \right)dtdh} \right)}, $$
where $s \ge 0$. The following lemma establishes that $I_{\vec{\lambda}}$ is of a non-degenerate distribution.  
\begin{lem} \label{Lemma: Non-degenerate Distribution} 
For all $ s \ge 0$, ${\int_0^\infty  {\int_0^\infty  {\left( {1 - {e^{ - s{{P}_{k}}hG_k\left( t \right)}}} \right)} {{\mu}_{k}}\left( t \right){q_{k}}\left( h \right)dtdh}}{< \infty}$.
\end{lem} 
\begin{IEEEproof}
Recall that $G_k(t) = \BO{t^{-\alpha_k}}$ as $t \ra \infty$.  Hence, we can find constants $B_1 > 0$ and ${\beta}>0$ such that $G_k(t) \leq {\beta} t^{-\alpha_k}$ for all $t \geq B_1$.  This implies that 
\begin{align}
&\int_0^\infty  {\int_0^\infty {\left( {1 - {e^{ - s{{P}_{k}}hG_k\left( t \right)}}} \right){{\mu}_{k}}\left( t \right){q_{k}}\left( h \right)dtdh} } \nonumber  \\
&  \le \int_0^\infty  {\int_0^{{B_1}} {\left( {1 - {e^{ - s{{P}_{k}}hG_k\left( t \right)}}} \right){{\mu}_{k}}\left( t \right){q_{k}}\left( h \right)dtdh + } } \int_0^\infty  {\int_{{B_1}}^\infty  {\left( {1 - {e^{ - s{{P}_{k}}h{\beta}{t^{ - \alpha_k }}}}} \right){{\mu}_{k}}\left( t \right){q_{k}}\left( h \right)dtdh} }  \nonumber   \\
&\le \int_0^{B_1}  {{{\mu}_{k}}\left( t \right)} {\kern 1pt} dt + \int_0^\infty  {\int_{{B_1}}^\infty  {\left( {1 - {e^{ - s{ {P}_{k} }h{\beta}{t^{ - \alpha_k }}}}} \right){{\mu}_{k}}\left( t \right){q_{k}}\left( h \right)dtdh}. }      \label{eq3}
\end{align}

The first integral in the last line in \eqref{eq3} is finite since ${\Lambda ^{\left( k \right)}}$ is locally finite. To show the finiteness of the second integral, we divide it into two parts as follows:
\begin{align}
&\int_0^\infty  {\int_{{B_1}}^\infty  {\left( {1 - {e^{ - s{{P}_{k}}h{\beta}{t^{ - \alpha_k }}}}} \right){{\mu}_{k}}\left( t \right){q_{k}}\left( h \right)dtdh} } \nonumber   \\
&= \int_0^{\frac{1}{{s{{P}_{k}}{\beta}}}} {\int_{{B_1}}^\infty  {\left( {1 - {e^{ - s{{P}_{k}}h{\beta}{t^{ - \alpha_k }}}}} \right){{\mu}_{k}}\left( t \right){q_{k}}\left( h \right)dtdh} }  \nonumber \\   
&+ \int_{\frac{1}{{s{{P}_{k}}{\beta}}}}^\infty  {\int_{{B_1}}^\infty  {\left( {1 - {e^{ - s{{P}_{k}}h{\beta}{t^{ - \alpha_k }}}}} \right){{\mu}_{k}}\left( t \right){q_{k}}\left( h \right)dtdh}. }   \label{eq4}
\end{align}
The first integral in (\ref{eq4}) can be bounded as 
\[\int_0^{\frac{1}{{s{{P}_{k}}{\beta}}}} {\int_{{B_1}}^\infty  {\left( {1 - {e^{ - s{{P}_{k}}h{\beta}{t^{ - \alpha_k }}}}} \right){{\mu}_{k}}\left( t \right){q_{k}}\left( h \right)dtdh} }  \le \int_{{B_1}}^\infty  {\left( {1 - {e^{ - {t^{ - \alpha_k }}}}} \right)} {{\mu}_{k}}\left( t \right)dt, \]
which is finite since $1 - {e^{ - {t^{ - \alpha_k }}}} = \BO{ {{t^{ - \alpha_k }}} }$ and ${{\mu}_{k}}\left( t \right) = \BO{ {{t^{\alpha_k  - 1 - \varepsilon }}} }$ as $t \to \infty$. Hence, proving the finiteness of $\int_{\frac{1}{{s{{P}_{k}}{\beta}}}}^\infty  {\int_{{B_1}}^\infty  {\left( {1 - {e^{ - s{{P}_{k}}h{\beta}{t^{ - \alpha_k }}}}} \right){{\mu}_{k}}\left( t \right){q_{k}}\left( h \right)dtdh} }$ will complete the proof. To this end, we need the following lemma.
\begin{lem}{\label{Lemma: Non-degenerate Distribution Supplementary} } 
$1 - {e^{ - a{t^{ - \alpha_k }}}} \le 2a\left( {1 - {e^{ - a}}} \right){t^{ - \alpha_k }}$ for all $a \ge 1$ and $t$ large enough.
\end{lem} 
\begin{IEEEproof} 
We let ${f_t}\left( a \right) = 1 - {e^{ - a{t^{ - \alpha_k }}}}$ and ${g_t}\left( a \right) = 2a\left( {1 - {e^{ - a}}} \right){t^{ - \alpha_k }}$. For $a=1$, we have $\mathop {\lim }\limits_{t \to \infty } \frac{{{f_t}\left( 1 \right)}}{{{t^{ - \alpha_k }}}} = 1$ and $\mathop {\lim }\limits_{t \to \infty } \frac{{{g_t}\left( 1 \right)}}{{{t^{ - \alpha_k }}}} = 2\left( {1 - {e^{ - 1}}} \right) > 1$. Hence, there exists a constant ${B_2} > 0$ such that ${g_t}\left( 1 \right) > {f_t}\left( 1 \right)$ for all $t \ge {B_2}$.  We now fix an arbitrary $t$ greater than ${B_2}$. Then, 
\begin{equation*}
\frac{{d{f_t}\left( a \right)}}{{da}} = {t^{ - \alpha_k }}{e^{ - a{t^{ - \alpha_k }}}}\quad \mbox{and } \quad \frac{{d{g_t}\left( a \right)}}{{da}} = 2{t^{ - \alpha_k }}\left( {1 + a{e^{ - a}} - {e^{ - a}}} \right).
\end{equation*}

Thus, ${{g_t}\left( a \right)}$ grows faster than ${{f_t}\left( a \right)}$, implying that ${{g_t}\left( a \right)} \ge {{f_t}\left( a \right)}$ for all $a \ge 1$ and $t \ge {B}_2$.
\end{IEEEproof}

By using Lemma \ref{Lemma: Non-degenerate Distribution Supplementary}, we can upper bound the second integral in (\ref{eq4}) as 
\begin{align}
&\int_{\frac{1}{{s{{P}_{k}}{\beta}}}}^\infty  {\int_{{B_1}}^\infty  {\left( {1 - {e^{ - s{{P}_{k}}h{\beta}{t^{ - \alpha_k }}}}} \right){{\mu}_{k}}\left( t \right){q_{k}}\left( h \right)dtdh} } \nonumber   \\
&\le \int_{{B_1}}^{{B_3}} {{{\mu}_{k}}\left( t \right)dt + \int_{{B_3}}^\infty  {\int_{\frac{1}{{s{{P}_{k}}{\beta}}}}^\infty  {2s{{P}_{k}}h{\beta}\left( {1 - {e^{ - s{{P}_{k}}h{\beta}}}} \right)} } } {q_{k}}\left( h \right){t^{ - \alpha_k }}{\mu}_{k}\left( t \right)dhdt \label{eq5}
\end{align}
for some positive constant $B_3$ large enough. The first integral in (\ref{eq5}) is finite due to local finiteness of $\Lambda^{(k)}$. The second integral in (\ref{eq5}) can be upper bounded by $2s{{P}_{k}}{\beta}{m}_{H}^{\left( k \right)}\int_{{B_3}}^\infty  {{t^{ - \alpha_k }}{{\mu}_{k}}\left( t \right)dt}$, which is finite since $m_{H}^{(k)} < \infty$ and ${{\mu}_{k}}\left( t \right) = \BO{ {{t^{\alpha_k  - 1 - \varepsilon }}} }$ as $t \to \infty$.
This completes the proof of Lemma \ref{Lemma: Non-degenerate Distribution}.
\end{IEEEproof}

The following lemma shows that the probability distribution of  $I_{\vec{\lambda}}$ can be approximated by the limit distribution of a sequence of random variables $I_n$, {\em i.e.,} ${I_n}\mathop  \to \limits^{\rm d} {I_{\vec{\lambda}}}$ as $n \to \infty$.
\begin{lem}  \label{Lemma: Distribution Convergence}  
For each $n$, let $U_{1,n}^{\left( k \right)}, \ldots ,U_{\left\lceil {\Lambda _n^{\left( k \right)}} \right\rceil ,n}^{\left( k \right)}$ be a sequence of i.i.d. random variables with a common probability density function ${f_{k}}\left( t \right) = \frac{{{\lambda _k}{{\mu}_{k}}\left( t \right)}}{{\Lambda _n^{\left( k \right)}}} \Inb{\left\{ {0 \le t \le n} \right\}}$ for tier-$k$, where $\Lambda _n^{\left( k \right)} = {\lambda _k}\int_0^n {{{\mu}_{k}}\left( t \right)} {\kern 1pt} dt$ and $\left\lceil . \right\rceil $ is the smallest integer greater than or equal to its argument. Let
\begin{eqnarray}
{I_n} = \sum\limits_{k = 1}^K {I_n^{\left( k \right)}}, \label{eq6}
\end{eqnarray}
where $I_n^{\left( k \right)} = P_k \sum_{i = 1}^{\left\lceil {\Lambda _n^{\left( k \right)}} \right\rceil } {H_{i}^{(k)}G_k\left( {U_{i,n}^{\left( k \right)}} \right)}$ and $\brparen{H_i^{(k)}}_{i=1}^\infty$ is an i.i.d. collection of random variables with the common probability density function $q_k(h)$ for $k=1, \ldots, K$.  Then $I_n$ converges in distribution to $I_{\vec{\lambda}}$, which is shown as ${I_n}\mathop  \to \limits^{\rm d} I_{\vec{\lambda}}$, as $n \to \infty $.
\end{lem} 
\begin{IEEEproof}
It is enough to show that $\mathcal{L}_{{I}_{n}}{(s)}$ converges to $\mathcal{L}_{{I}_{\vec{\lambda}}}{(s)}$ pointwise as $n$ tends to infinity.  Observing that the random variables $I_n^{(k)}$ for $k=1, \ldots, K$ are independent, we can write the Laplace transform of $I_n$ as  
\begin{eqnarray}
\mathcal{L}_{{I}_{n}}\left( s \right) = \prod_{k=1}^K\ES{\e{-s I_n^{(k)}}} = \prod_{k=1}^K \mathcal{L}_{I_n^{(k)}}(s), \nonumber
\end{eqnarray}
where ${{{\mathcal{L}}_{I_n^{\left( k \right)}}}\left( s \right)}$ is the Laplace transform of $I_n^{(k)}$, which is given by
\begin{equation*}
{{\mathcal{L}}_{I_n^{\left( k \right)}}}\left( s \right) =
{\left( {1 - \frac{{{\lambda _k}}}{{\Lambda _n^{\left( k \right)}}}\int_0^\infty  {\int_0^n {\left( {1 - {e^{ - s{{P}_{k}}hG_k\left( t \right)}}} \right)} {{\mu}_{k}}\left( t \right){q_{k}}\left( h \right)dtdh} } \right)^{\left\lceil {\Lambda _n^{\left( k \right)}} \right\rceil }}. 
\end{equation*}

As $n$ grows to infinity, $\int_0^\infty  {\int_0^n {\left( {1 - {e^{ - s{{P}_{k}}hG_k\left( t \right)}}} \right)} {{\mu}_{k}}\left( t \right){q_{k}}\left( h \right)dtdh}$ converges to 
$$\int_0^\infty  {\int_0^\infty  {\left( {1 - {e^{ - s{{P}_{k}}hG_k\left( t \right)}}} \right)} {{\mu}_{k}}\left( t \right){q_{k}}\left( h \right)dtdh}$$ 
and $\int_0^\infty  {\int_0^\infty  {\left( {1 - {e^{ - s{{P}_{k}}hG_k\left( t \right)}}} \right)} {{\mu}_{k}}\left( t \right){q_{k}}\left( h \right)dtdh}  < \infty$ by Lemma \ref{Lemma: Non-degenerate Distribution}.  This observation leads to the following identity 
\begin{equation*}
\mathop {\lim}\limits_{n \to \infty } {{\mathcal{L}}_{I_n^{\left( k \right)}}}\left( s \right) = \exp \left( { - {\lambda _k}\int_0^\infty  {\int_0^\infty  {\left( {1 - {e^{ - s{{P}_{k}}hG_k\left( t \right)}}} \right)} {{\mu}_{k}}\left( t \right){q_{k}}\left( h \right)dtdh} } \right), 
\end{equation*}
which is exactly the Laplace transform of the AWI at the test user coming from tier-$k$ BSs alone.  Utilizing this result, we have
\begin{eqnarray}
\lim_{n \ra \infty} \mathcal{L}_{I_n}(s) &=& \lim_{n \ra \infty} \prod_{k=1}^K \mathcal{L}_{I_n^{(k)}}(s) \nonumber  \\ 
&=& \prod_{k=1}^K \lim_{n \ra \infty} \mathcal{L}_{I_n^{(k)}}(s) \nonumber \\
&=& \prod_{k=1}^K \exp \left( { - {\lambda _k}\int_0^\infty  {\int_0^\infty  {\left( {1 - {e^{ - s{{P}_{k}}hG_k\left( t \right)}}} \right)} {{\mu}_{k}}\left( t \right){q_{k}}\left( h \right)dtdh} } \right) \nonumber \\
&=& \mathcal{L}_{I_{\vec{\lambda}}}(s), 
\end{eqnarray}
which completes the proof. 
\end{IEEEproof}

The next lemma shows that the mean value and variance of $I_{\vec{\lambda}}$ can also be approximated by the mean value and variance of $I_n$. 
\begin{lem} \label{Lemma: Mean Variance Convergence} 
Let $I_{n}$ be defined as in (\ref{eq6}). Then,
\[\lim\limits_{n \to \infty } {\EW}\left[ {{I_n}} \right] = {\EW}\left[ {{I_{\vec{\lambda}} }} \right]\]
and
\[\lim\limits_{n \to \infty } {\V{ {I_n} }} = {\V{ {I_{\vec{\lambda}}} }} \]
\end{lem} 
\begin{IEEEproof} 
Using Campbell's Theorem \cite{Kingman93}, we can express $\ES{I_{\vec{\lambda}}}$ and $\V{I_{\vec{\lambda}}}$ as
$$ \ES{I_{\vec{\lambda}}} = \sum_{k=1}^K \lambda_k P_k m_H^{(k)} \int_0^\infty G_k(t) \mu_k(t) dt$$
and 
$$\V{I_{\vec{\lambda}}} = \sum_{k=1}^K \lambda_k P_k^2 m_{H^2}^{(k)} \int_0^\infty G_k^2(t) \mu_k(t) dt. $$

We note that our modeling assumptions ensure that ${{\EW}\left[ {{I_{{\lambda _k}}}} \right]}$ and ${\V{ {I_{\lambda_k} } }}$ are both finite numbers. Let the random variables $U_{i,n}^{\left( k \right)}$, $H_i^{(k)}$ and $I_n^{(k)}$ be as defined in Lemma \ref{Lemma: Distribution Convergence}. Further, let $m_{i, n}^{(k)} = \ES{P_k H_i^{(k)} G_k\paren{U_{i, n}^{(k)}}}$ and $\sigma_{i, n}^{(k)} = \sqrt{\V{P_k H_i^{(k)} G_k\paren{U_{i, n}^{(k)}}}}$. We first observe that
$$ \ES{I_n^{(k)}} = \Ceil{\Lambda_n^{(k)}} m_{1, n}^{(k)} \hspace{0.5cm} \mbox{and} \hspace{0.5cm} \V{I_n^{(k)}} = \Ceil{\Lambda_n^{(k)}} \paren{\sigma_{1, n}^{(k)}}^2. $$
Furthermore, we can express $m_{i,n}^{\left( k \right)}$ as 
$$
m_{i,n}^{\left( k \right)} = \frac{{{\lambda _k}{{P}_{k}}m_H^{\left( k \right)}}}{{\Lambda _n^{\left( k \right)}}}\int_0^n {G_k\left( t \right){{\mu}_{k}}\left( t \right)dt}, 
$$ 
which implies that $\lim_{n \ra \infty} \ES{I_n^{(k)}} = \lambda_k P_k m_H^{(k)} \int_0^\infty G_k(t) \mu_k(t) dt$.  Using this result, we have
\begin{eqnarray}
\lim_{n \ra \infty} \ES{I_n} &=& \lim_{n \ra \infty} \sum_{k=1}^K \ES{I_n^{(k)}} \nonumber \\
&=& \sum_{k=1}^K \lim_{n \ra \infty} \ES{I_n^{(k)}} \nonumber \\
&=& \sum_{k=1}^K \lambda_k P_k m_H^{(k)} \int_0^\infty G_k(t) \mu_k(t) dt \nonumber \\
&=& \ES{I_{\vec{\lambda}}}. \nonumber
\end{eqnarray}

Repeating the similar steps and using the identity 
\begin{equation*}
{\left( {\sigma _{i,n}^{\left( k \right)}} \right)^2} = \frac{{{\lambda _k}{{ {{{P}_{k}^2}}}}m_{{H^2}}^{\left( k \right)}}}{{\Lambda _n^{\left( k \right)}}}\int_0^n {{G_k^2}\left( t \right){{\mu}_{k}}\left( t \right)dt} - \frac{{\lambda _k^2{{ {{{P}_{k}^2}} }}{{\left( {m_H^{\left( k \right)}} \right)}^2}}}{{{{\left( {\Lambda _n^{\left( k \right)}} \right)}^2}}}{\left( {\int_0^n {G_k\left( t \right){{\mu}_{k}}\left( t \right)dt} } \right)^2}, 
\end{equation*}
we also obtain $\lim_{n \ra \infty}\V{I_n} = \V{I_{\vec{\lambda}}}$. 
\end{IEEEproof}

\begin{lem} \label{Lemma: Berry Esseen} 
Let ${\xi _1}, \ldots ,{\xi _m}$ be a sequence of independent and real-valued random variables such that ${\EW}\left[ {{\xi _i}} \right] = 0$ and $\sum\nolimits_{i = 1}^m {{\EW}\left[ {\xi _i^2} \right]}  = 1$. Let $\chi  = \sum\nolimits_{i = 1}^m {{\EW}\left[ {\left| {\xi _i^3} \right|} \right]}$. Then,
$$
\left| \PR{ {\sum_{i = 1}^m {{\xi _i} \le x} }} - \Psi(x) \right| \le \chi \min \left( {0.4785,\frac{{31.935}}{{1 + {{\left| x \right|}^3}}}} \right)
$$ 
for all $x \in \R$.
\end{lem}
\begin{IEEEproof}
Please refer to \cite{Inaltekin12}.   
\end{IEEEproof}

\section{Proof of Theorem \ref{Theorem 1}} \label{Appendix:Theorem_1}

In this appendix, we provide the proof for our main Gaussian approximation result given in Theorem \ref{Theorem 1}.  To this end, we let $\xi _{i,n}^{\left( k \right)} = \frac{{{{P}_{k}}H_i^{\left( k \right)}G_k\left( {U_{i,n}^{\left( k \right)}} \right) - m_{i,n}^{\left( k \right)}}}{{{\sigma _n}}}$ for $k=1, \ldots, K$, $n \ge 1$ and $1 \le i \le \left\lceil {\Lambda _n^{\left( k \right)}} \right\rceil$, where ${\sigma _n} = \sqrt { \V{I_n} }$, and $I_n$, ${U_{i,n}^{\left( k \right)}}$, ${\Lambda _n^{\left( k \right)}}$ and $m_{i,n}^{\left( k \right)}$ are as defined in Appendix \ref{Preliminary}. We note that $\ES{\xi _{i,n}^{\left( k \right)}} = 0$ and $\sum_{k=1}^K \sum_{i=1}^{\left\lceil {\Lambda _n^{\left( k \right)}} \right\rceil} \ES{\paren{\xi _{i,n}^{\left( k \right)}}^2} = 1$.  Hence, the collection of random variables $\bigcup_{k=1}^K\brparen{\xi _{i,n}^{\left( k \right)}: i=1, \ldots, \left\lceil {\Lambda _n^{\left( k \right)}} \right\rceil}$ is in the correct form to apply Lemma \ref{Lemma: Berry Esseen}. We need to calculate $\chi_n = \sum_{k=1}^K \sum_{i=1}^{\left\lceil {\Lambda _n^{\left( k \right)}} \right\rceil} \abs{\xi_{i, n}^{(k)}}^3$ to complete the proof.  We can upper bound $\chi_n$ as     
\begin{eqnarray*}
\lefteqn{\chi_n \leq \frac{1}{\sigma_n^3} \sum_{k=1}^K \Ceil{\Lambda_n^{(k)}} \ES{\abs{P_k H_1^{(k)} G_k\paren{U_{1, n}^{(k)}} + m_{1, n}^{(k)}}^3}} \hspace{16cm} \\
\lefteqn{= \frac{1}{\sigma_n^3} \sum_{k=1}^K \Ceil{\Lambda_n^{(k)}} \EW\left[P_k^3 \paren{H_1^{(k)}}^3 \paren{G_k\paren{U_{1, n}^{(k)}}}^3 + 3P_k^2 \paren{H_1^{(k)}}^2 \paren{G_k\paren{U_{1, n}^{(k)}}}^2 m_{1, n}^{(k)} \right.} \hspace{15.3cm} \\
\lefteqn{ \left. + 3P_k H_1^{(k)} G_k\paren{U_{1, n}^{(k)}} \paren{m_{1, n}^{(k)}}^2 + \paren{m_{1, n}^{(k)}}^3 \right]} \hspace{6.8cm} \\
\lefteqn{= \frac{1}{\sigma_n^3} \sum_{k=1}^K \Ceil{\Lambda_n^{(k)}} \left(\frac{P_k^3 m_{H^3}^{(k)} \lambda_k}{\Lambda_n^{(k)}} \int_0^n \paren{G_k(t)}^3 \mu_k(t) dt + 3\frac{P_k^2 m_{H^2}^{(k)} \lambda_k}{\Lambda_n^{(k)}} \int_0^n \paren{G_k(t)}^2 \mu_k(t) dt \cdot m_{1, n}^{(k)}\right.} \hspace{15.3cm} \\
\lefteqn{\left. +3\frac{P_k m_{H}^{(k)} \lambda_k}{\Lambda_n^{(k)}} \int_0^n G_k(t) \mu_k(t) dt \cdot \paren{m_{1, n}^{(k)}}^2 + \paren{m_{1, n}^{(k)}}^3 \right).} \hspace{9cm} \\
\end{eqnarray*}

We note that $m_{1,n}^{\left( k \right)} = o\left( 1 \right)$ and $\Ceil{\Lambda_n^{(k)}} \paren{m_{1,n}^{\left( k \right)}}^3 = o\left( 1 \right)$ as $n \to \infty$, {\em i.e.}, see the proof of Lemma \ref{Lemma: Mean Variance Convergence}. Furthermore, we know that $\sigma _n^2$ converges to $\V{I_{\vec{\lambda}}}$ as $n \to \infty$ by Lemma \ref{Lemma: Mean Variance Convergence}.  Using these results, we have
\begin{eqnarray}
\limsup_{n \ra \infty} \chi_n \leq \frac{1}{\paren{\V{I_{\vec{\lambda}}}}^\frac32} \sum_{k=1}^K P_k^3 m_{H^3}^{(k)} \lambda_k \int_0^\infty \paren{G_k(t)}^3 \mu_k(t) dt. \label{eq8}
\end{eqnarray}
After substituting the expression for $\V{I_{\vec{\lambda}}}$ (see the proof of Lemma \ref{Lemma: Mean Variance Convergence}) in \eqref{eq8}, we obtain
\begin{eqnarray}
\limsup_{n \ra \infty} \chi_n \leq \sum_{k=1}^K \frac{\lambda_k P_k^3 m_{H^3}^{(k)} \int_0^\infty G_k^3(t) \mu_k(t) dt}{\paren{\sum_{k=1}^K \lambda_k P_k^2 m_{H^2}^{(k)} \int_0^\infty G_k^2(t) \mu_k(t) dt}^\frac32}. \label{eq9}
\end{eqnarray} 
By using Lemma \ref{Lemma: Berry Esseen}, we have
\begin{eqnarray} 
\abs{\PR{\sum_{k=1}^K \sum_{i=1}^{\Ceil{\Lambda_n^{(k)}}} \xi_{i, n}^{(k)} \leq x} - \Psi(x)} \leq \chi_n \min\paren{0.4785, \frac{31.935}{1+\abs{x}^3}} \label{eq10} 
\end{eqnarray}
for all $n \geq 1$ and $x \in \R$.  Further, Lemmas \ref{Lemma: Distribution Convergence} and \ref{Lemma: Mean Variance Convergence} imply that
\begin{eqnarray}
\sum_{k=1}^K\sum_{i=1}^{\Ceil{\Lambda_n^{(k)}}} \xi_{i, n}^{(k)} \mathop  \to \limits^{\rm d} \frac{I_{\vec{\lambda}} - \ES{I_{\vec{\lambda}}}}{\sqrt{\V{I_{\vec{\lambda}}}}} \quad \mbox{ as } \quad n \ra \infty. \label{eq11} 
\end{eqnarray}
Hence, using \eqref{eq9} and taking the $\limsup$ of both sides in \ref{eq10}, we have
\begin{eqnarray*}
\lefteqn{\limsup_{n \ra \infty} \abs{\PR{\sum_{k=1}^K \sum_{i=1}^{\Ceil{\Lambda_n^{(k)}}} \xi_{i, n}^{(k)} \leq x} - \Psi(x)}} \hspace{16cm} \\
\lefteqn{ = \abs{\PR{\frac{I_{\vec{\lambda}} - \ES{I_{\vec{\lambda}}}}{\sqrt{\V{I_{\vec{\lambda}}}}} \leq x} - \Psi(x)}} \hspace{11.5cm} \\
\lefteqn{\leq \sum_{k=1}^K \frac{\lambda_k P_k^3 m_{H^3}^{(k)} \int_0^\infty G_k^3(t) \mu_k(t) dt}{\paren{\sum_{k=1}^K \lambda_k P_k^2 m_{H^2}^{(k)} \int_0^\infty G_k^2(t) \mu_k(t) dt}^\frac32} \min\paren{0.4785, \frac{31.935}{1+\abs{x}^3}},} \hspace{11.5cm}
\end{eqnarray*}
which completes the proof.


\ifCLASSOPTIONcaptionsoff
  \newpage
\fi



%

%

\vfill \vfill
\end{document}